\documentclass[twocolumn]{aastex631}
\usepackage{etoolbox}
\makeatletter
\patchcmd{\ltx@foottext}{%
  .5\textwidth\advance\hsize-18pt}{%
  \linewidth\advance\hsize-1.8em%
}{}{}
\makeatother
\shorttitle{Transverse MHD waves as signatures of braiding-induced magnetic reconnection in coronal loops}
\shortauthors{Sukarmadji \& Antolin.}

\graphicspath{{./}{figures/}}

\begin{document}

\title{Transverse MHD waves as signatures of braiding-induced magnetic reconnection in coronal loops}

\author[0000-0002-9723-3155]{A. Ramada C. Sukarmadji}
\affiliation{Department of Mathematics, Physics and Electrical Engineering, Northumbria University \\ Newcastle upon Tyne, NE1 8ST, UK}

\author[0000-0003-1529-4681]{Patrick Antolin}
\affiliation{Department of Mathematics, Physics and Electrical Engineering, Northumbria University \\ Newcastle upon Tyne, NE1 8ST, UK}

\begin{abstract}

A major coronal heating theory based on magnetic reconnection relies on the existence of braided magnetic field structures in the corona. In this small-angle reconnection scenario, numerical simulations indicate that the reconnected magnetic field lines are driven sideways by magnetic tension and can overshoot from their new rest position, thereby leading to low-amplitude transverse MHD waves. This provides an efficient mechanism for transverse MHD wave generation, and the direct causality also constitutes substantial evidence of reconnection from braiding. However, this wave-generation mechanism has never been directly observed. Recently, the telltale signature of small-angle reconnection in a sheared coronal structure has been identified through nanojets, which are small, short-lived, and fast jet-like bursts in the nanoflare range transverse to the guide-field. We present for the first time IRIS and SDO observations of transverse MHD waves in a coronal loop that directly result from braiding-induced reconnection. The reconnection is identified by the presence of nanojets at the loop apex which release nanoflare-range energy. We find that the oscillations have an energy flux on the order of $10^6 - 10^8$~erg~cm$^{-2}$~s$^{-1}$, which is within the budget to power active region loops. The estimated kinetic and thermal energy from the nanojets is also sufficient to power the transverse waves and sustain the observed heating at the loop apex.  This discovery provides major support to (a) existing theories that transverse MHD waves can be a signature of reconnection, (b) the existence of braiding in coronal structures and (c) the coronal reconnection scenario identified by nanojets.

\end{abstract}


\section{Introduction} \label{sec:intro}

Magnetohydrodynamic (MHD) waves and magnetic reconnection are the two leading theories for the coronal heating problem, which has been a subject of investigation for decades. MHD waves can carry large amounts of energy \citep{uchida1974}, making them a suitable candidate for heating through their dissipation (\citealt{wentzel1979}; \citealt{klimchuk2006}; \citealt{vandoorsselaere2020}). This was supported by the discovery of transverse oscillations in loop-like structures (\citealt{aschwanden1999}; \citealt{nakariakov1999}), and followed up by many other works reporting similar oscillations (\citealt{aschwanden2002}; \citealt{verwichte2004}; \citealt{tomczyk2007}; \citealt{tomczyk2009}; \citealt{vandoorsselaere2009}; \citealt{okamoto2015}; \citealt{anfinogentov2015}; \citealt{li2023}).  However, direct observational evidence of wave-based heating remains scarce \citep{vandoorsselaere2020}.

Among transverse oscillations, the kink oscillations \citep{nakariakov2021} are the most prominently observed and often found to be decayless (\citealt{tian2012}; \citealt{wang2012}; \citealt{nistico2013}; \citealt{anfinogentov2013}). Many external driving mechanisms have been proposed for these decayless oscillations, including footpoint driving (\citealt{nistico2013}), quasi-steady flows (\citealt{nakariakov2016}),  or by Alfv\'enic vortex shedding (\citealt{karampelas2021}). Proposed internal mechanisms include the combination of resonant absorption and the Kelvin-Helmholtz instability (\citealt{antolin2016}; \citealt{antolin2019}) or coronal rain (\citealt{kohutova2017}), but there is still a lack of observational evidence.

Another major coronal heating candidate is the Parker nanoflare theory \citep{parker1988}, which conjectures the existence of myriad energy bursts at the order of $10^{24}$~erg generated by small-scale magnetic reconnection events driven by braiding. The braided state of a loop is thought to be the result of slow footpoint motions at photospheric level \citep{pontin2022}. Energy releases within the nanoflare range have been previously reported by e.g. \cite{testa2013} and \cite{testa2014}, and \cite{chitta2018} have also suggested that chromospheric reconnection from flux cancellation in loop footpoints may facilitate nanoflare-sized energy release in loops. However, a direct link to coronal reconnection could not be established in these heating events.

The discovery of nanojets by \cite{antolin2020} provided direct evidence of nanoflare-based heating driven by small-scale component reconnection. Nanojets are small-scale and short-lived bursts, around 500~km in width and 1000~km in length on average, that last no longer than 15~s on average. They are a result of very fast transverse motion of reconnected field lines driven by magnetic tension, combined with localised heating (the nanoflare). In \cite{antolin2020}, they were observed in a loop-like structure and driven by the loss of stability of a nearby prominence. This was then followed by observations of nanojets in loop-like structures with coronal rain \cite{sukarmadji2022}, with the Kelvin-Helmholtz instability (KHI) and Rayleigh-Taylor instability (RTI) as the likely underlying drivers. The different observations of nanojets in a variety of environments with different drivers further suggests that they may be common, and could contribute significantly to the heating of the solar corona.

It has been long known that magnetic reconnection can produce all kinds of MHD waves (e.g. \citealt{petschek1964}, \citealt{parker1991}, \citealt{kigure2010}), however, it is unclear which waves would be predominantly observed in the Parker nanoflare theory. The kink instability as a trigger of reconnection and driver of coronal heating has been extensively studied through numerical simulations of twisted magnetic fields (\citealt{browning2008}; \citealt{hood2009}; \citealt{bareford2013}; \citealt{hood2016}; \citealt{reid2018}; \citealt{reid2020}). In particular, \cite{hood2009}, \cite{hood2016}, \cite{reid2018}, and \cite{reid2020} proposed the existence of twisted coronal braids or strands, some of which would become unstable thereby setting a cascade of nanoflare-sized reconnection events affecting neighboring stable strands. Although not investigated in detail, these works show the generation of transverse MHD waves during the reconnection process. Observationally, \cite{kohutova2020} have detected torsional Alfv\'en waves produced from a reconnection event, although the configuration leading to the reconnection corresponds to the presence of 2 separate coronal structures and not a single braided structure. To date, there are no direct observations of small-angle reconnection events leading to kink waves. Yet, as mentioned previously, kink waves are ubiquitous in the solar corona and their origin is highly debated.

We present in this paper first observations of transverse oscillations driven by small-angle reconnection events in a coronal loop, where the reconnections are identified by the presence of nanojets. We will look into the properties of the waves produced, and discuss the heating contributed from the observed event. In Section \ref{sec:observations}, we present the observation and the present structures. Section \ref{sec:nanojets} discusses the reconnection nanojets, followed with a discussion of the transverse waves produced in Section \ref{sec:transverse}. We will discuss the energy budget in Section \ref{sec:energy} and provide conclusions in Section \ref{sec:discussion}.

\section{Observations} \label{sec:observations}

An observation of AR 12192 was taken by the Interface Region Imaging Spectrograph (IRIS; \citealt{depontieu2014}) on the 29th of October 2014 between 08:37:04-13:43:35 UT,  observing in the SJI 1330 filtergram with spatial resolution, temporal cadence, and exposure time of $0.16\arcsec$, 9.6~s, and 8~s respectively. This is a large coarse 8-step raster observation centered at (x,y) = ($956\arcsec$,$-262\arcsec$) with a field-of-view (FOV) of $119\arcsec \times 119\arcsec$. We use the level 2 data for our analysis, along with coaligned observations from the Atmospheric Imaging Assembly (AIA; \citealt{lemen2012}). During the observing period, the area produced a series of C to M-class flare events with a number of surges and quiescent and flaring coronal rain.

Our focus will be on a hot loop shown in Figure \ref{fig1}. We initially observe a diffuse bright loop-like region at 11:49:37~UT only visible in AIA 94 and 131 and faintly in the SJI 1330 (the Fe XXI emission), which disappears after 13:18:01~UT.  Following this disappearance, the loop is seen to form at 13:23:37 UT in SJI 1330, AIA 304, 171, 193, 211, and faintly in 131 and 335, suggesting catastrophic cooling and the appearance of coronal rain. The time range of interest starts at 13:30:19 UT, when the cool loop exhibits a secondary heating event and is seen in all changes but only faintly in AIA 94, 335 and 1600. The structure remains visible until the end of the observing period of IRIS, and until 14:19:07 UT in AIA. Unfortunately, the IRIS slit does not cross our loop of interest, and therefore we can only obtain plane-of-sky (POS) values for all of our measurements.

The loop appears first at a measured height at the apex of $29 \pm 5$~Mm as measured in the POS from the solar surface, and a length of $120 \pm 20$~Mm in the POS. Within the loop, we observe coronal rain flowing with POS velocities of 20-32~km~s$^{-1}$ at the apex and 100-120~km~s$^{-1}$ along the legs of the loop. The rain strands have widths of $600 \pm 45$~km, and the apparent width of the loop at the apex seen in the upper transition region or coronal AIA passbands is $4059 \pm 489$~km.

We used the Basis Pursuit Method, based on the Differential Emission Measure (DEM) Analysis (\citealt{cheung2015}) to estimate the emission distribution with respect to temperature. The DEM weighted electron number density of the loop $\langle n_e \rangle_{DEM}$ depends on the emission $DEM(T)$ in the LOS $l$ for a given temperature bin
 \begin{equation} \label{eq:1} 
DEM(T) = n_e n_H \frac{dl}{dT},  
\end{equation} 
where $n_e$ and $n_H$ are the electron and hydrogen number densities, respectively.  $\langle n_e \rangle_{DEM}$ follows 
 \begin{equation} \label{eq:2} 
\langle n_e \rangle_{DEM} = \sqrt{\frac{\int_T DEM(T)_{loop}dT}{1.2 l}},  
\end{equation} 
where we assume a fully ionised plasma with $10\%$ Helium abundance and $l$ is the length of the emitting plasma in the loop along the LOS. $DEM(T)_{loop} = DEM(T)_{LOS} - DEM(T)_{env}$ with $DEM(T)_{LOS}$ is the DEM at a given temperature bin for a LOS crossing the loop, and $DEM(T)_{env}$ is the DEM for a LOS neighboring the loop in the same temperature bin. We thus assume that the foreground and background along this neighboring LOS is the same as that crossing the loop so that the subtraction gives the DEM of the loop. We averaged the EM values from the DEM calculation that has converged, for pixels contained in the loop for each temperature bin. The temperature bins used are $log(T) = 5.5 - 6.4$ as they show emission from the loop.

We assume that $l$ is similar to the POS width of the loop to obtain an electron number density of $4.7 \pm 0.5 \times 10^{9}$~cm$^{-3}$, which corresponds to the optically thin hot plasma surrounding the coronal rain strands. The number density of the cool rain strands is estimated through pressure balance and taking the peak temperature response for SJI 1330 ($10^{4.3}$~K),  to find $3.8 \pm 0.8 \times 10^{11}$~cm$^{-3}$. This matches previous measurements of coronal rain densities in observations (\citealt{antolin2015b}; \citealt{froment2020}) and numerical simulations (\citealt{li2022}; \citealt{antolinms2022}) .

\section{Reconnection Nanojets} \label{sec:nanojets}

\begin{figure*}[ht]
\centering
\includegraphics[width=0.8\textwidth]{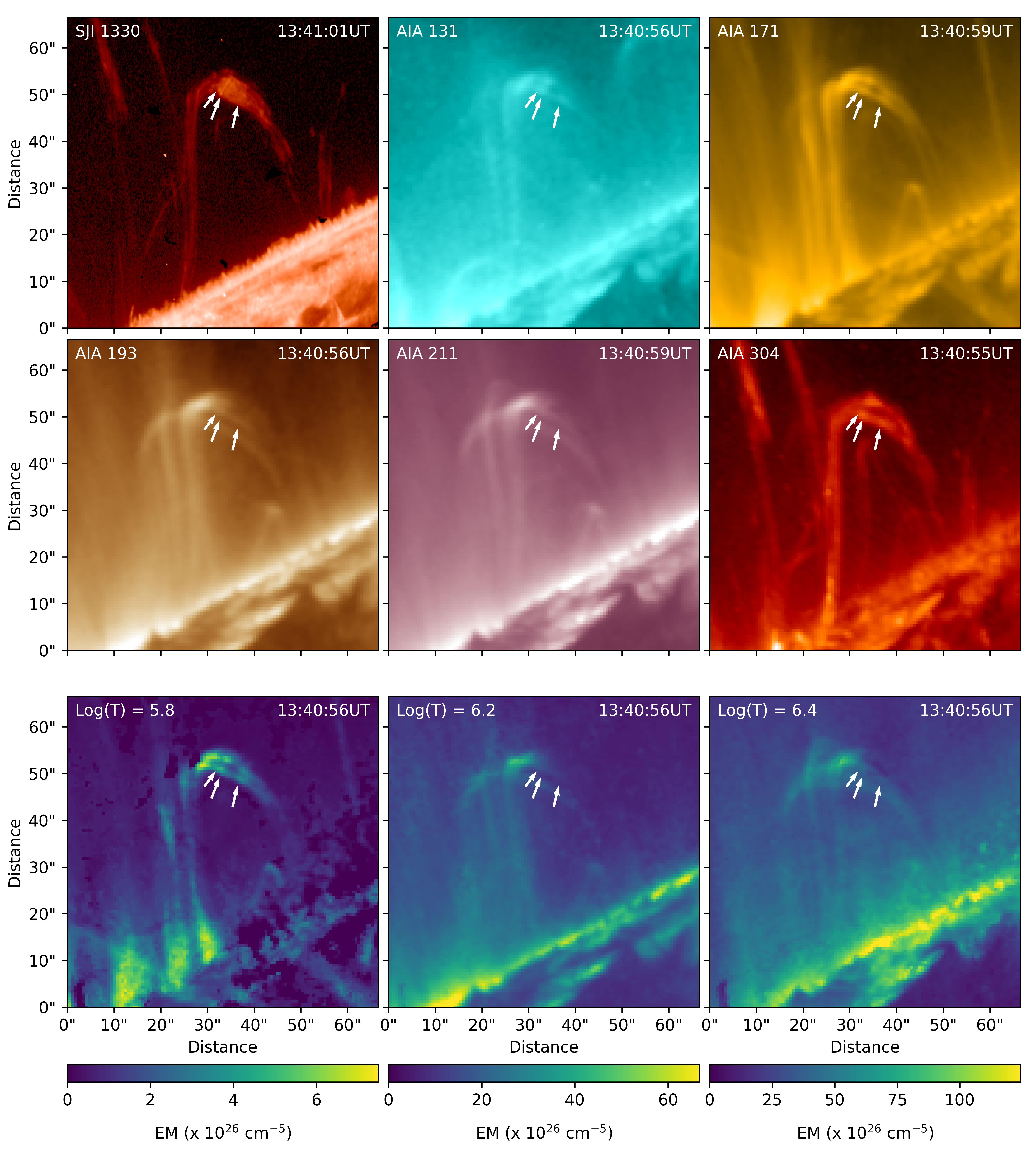}
\caption{First two rows shows the snapshots of the loop at the time when the nanojets are most visible in SJI 1330, AIA 131, 171, 193, 211, and 304. The bottom row shows snapshots for selected emission bins from Log(T) = 5.8, 6.2 and 6.4. Three of the clearest nanojets (left to right, N1, N2, and N3) are marked with the white arrows. An animation of this figure is available online showing the time evolution of the loop.
\label{fig1}}
\end{figure*}

\begin{figure*}[t!]
\centering
\includegraphics[width=0.8\textwidth]{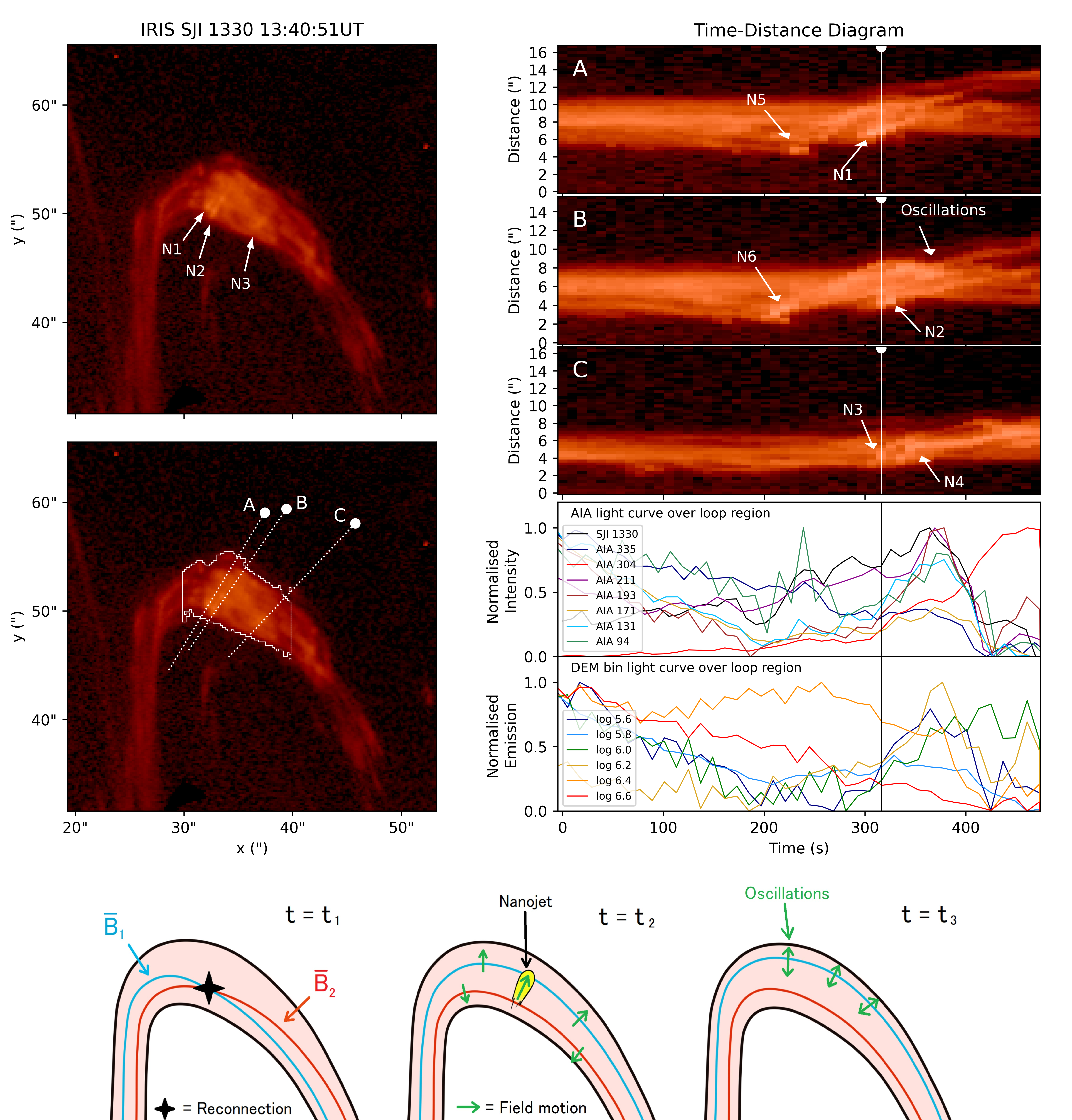}
\caption{The two panels of the left column show the same snapshot of the apex of the loop with nanojets. In the top panel, the three most visible nanojets are marked, and in the bottom panel, slices along the trajectory of the nanojets A, B, and C are taken to produce the time-distance diagrams on the first three panels of the right column. The nanojets in the time-distance diagrams are marked with N1-N6. The white vertical lines in the time-distance diagrams mark the time of the snapshot from the left column. The region contained by the white contour line in the bottom panel of the left row shows the region used to produce the light-curve plot shown in the bottom panel of the right column. The light curves are constructed by summing over the intensity values within the contour at a given timestamp and then normalised. In the bottom we show a schematic for our interpretation of how a nanojet forms from reconnection due to misalignments between $\vec{B}_1$ and $\vec{B}_2$. The resulting configuration is likely to be affected by the tension from the reconnection, which overshoots the resting configuration and therefore produces oscillations, as seen at $t = t_3$. An animation of this figure without the schematic is available online, showing the nanojet formation in the SJI, and the white vertical line in the time distance diagrams and light curves following the timestamp of the SJI image. The 7 other observed nanojets are also marked in the animation with white arrows (nanojets pointing upwards) and cyan arrows (nanojets pointing downwards).
\label{fig2}}
\end{figure*}

The reconnection is identified by the presence of nanojets at the apex of the loop, shown in Figure \ref{fig1} and the top left panel of Figure \ref{fig2}. After 13:35:35 UT, we begin to observe around 10 small jet-like structures characteristic of nanojets forming at the loop apex with 7 of them oriented upwards (away from the solar surface) and the remaining oriented downwards (towards the solar surface). Following \cite{antolin2020} and \cite{sukarmadji2022}, we identify these structures as nanojets if they have the small, transverse (to the loop) jet-like features in the image and the running difference image (current image subtracted with image from previous timestamp) with widths and lengths of around 500~km and 1000~km, respectively, and are accompanied with a transverse motion fairly perpendicular to the loop and with short timescales of less than $25$~s. We investigated 3 of the clearest ones (N1, N2, and N3) to have a measure of their properties. The mean of their POS lengths and widths are 1500~km and 341~km, with values ranging from $838 - 2781$~km and $320 - 405$~km respectively. The lifetimes of all three nanojets are $19 \pm 5$~s. We also measured the DEM weighted electron number densities and temperatures of the nanojets. The electron number density is measured through Equation \ref{eq:2}, using the emission from the pixel containing the nanojet and assuming that $l$ is similar to the POS width of the nanojet. The nanojet's temperature is measured through
\begin{equation} \label{eq:3}
T = T_{t_0} + \Delta T,
\end{equation}
averaged over all the nanojet pixels, where $T_{t_{0}}$ is the DEM weighted temperature of the region before the nanojet forms, following 
\begin{equation} \label{eq:4}
T_{t_0} = \frac{\int DEM(T)_{t_0}TdT}{\int DEM(T)_{t_0}dT},
\end{equation}
and $\Delta T$ is the temperature change at the nanojet timestamp, measured by the variation of the DEM: 
\begin{equation} \label{eq:5}
\Delta T = \frac{\int \Delta DEM(T)TdT}{\int\Delta DEM(T)dT},
\end{equation}
where $\Delta DEM(T) = DEM(T)_{nanojet} - DEM(T)_{t_0}$. $DEM(T)_{nanojet}$ is the DEM at the nanojet timestamp and $DEM(T)_{t0}$ is the DEM at a timestamp before the nanojet forms. We find a mean number density and temperature of  $1.2 \times 10^{10}$~cm$^{-3}$ and 2.3~MK, with values ranging from $0.9 - 1.4 \times 10^{10}$~cm$^{-3}$ and $2.2 - 2.4$~MK, respectively. The nanojets have an average POS speed of 156~km~s$^{-1}$, ranging from $91-290$~km~s$^{-1}$.

To obtain a measure of the kinetic and thermal energy, the kinetic energy ($E_K$) and thermal energy ($E_T$) is calculated through $E_{K} = \frac{1}{2}1.27V\langle n_{e}\rangle_{DEM}m_{p}v^{2}$ and $E_{T} = \frac{3}{2}2.09V\langle n_{e}\rangle_{DEM}kT$, where $V$ is the nanojet volume, $m_{p}$ is the proton mass, $k$ is the Boltzmann constant, and $T$ is the average nanojet temperature (following Equation \ref{eq:3}). The factors 1.27 and 2.09 comes from assuming a 10\% Helium abundance and a highly ionized plasma \footnote{Although coronal rain corresponds to partially ionised plasma, numerical simulations show that its ionisation fraction is relatively high (\citealt{antolin2022b}).}. We assume that the nanojet has a cylindrical structure with radius and length set by the observed mean values, and $v$ is set equal to the mean POS velocity. This gives us energy releases within the nanoflare range, with a kinetic and thermal energy release average of $7.8 \times 10^{24}$~erg and $1.4 \times 10^{25}$~erg per nanojet, with values of $0.8 - 21.7 \times 10^{24}$~erg and $0.9 - 2.3 \times 10^{25}$~erg, respectively. The average total energy per nanojet is therefore $2.2 \times 10^{25}$~erg, and as we observe around 10 nanojets, the total estimated energy released from all nanojets is $2.2 \times 10^{26}$~erg. From these measured cases, the nanojets have similar morphologies, dynamics, and energy release as in \cite{antolin2020} and \cite{sukarmadji2022}, although slightly smaller in size. We also observe local brightenings in the hot AIA channels in the regions where we observe nanojets, as shown in in Figure \ref{fig1}, supporting the presence of localised heating in the loop-like structure. The nanojets are particularly seen in AIA 131, 171, 193, 211, 304, and the DEM bin Log(T) = 5.8 and 6.4.

These signatures are clearer in a light curve plot integrating over the loop region containing the nanojets, shown in Figure \ref{fig2} (see figure caption for methods), where the occurrence of nanojets coincides with peaks of the average loop apex intensity in various passbands (all except 335). The light curve intensity for the SJI 1330, AIA channels 211, 193, 171, 131 ramps up to two intensity peaks as the nanojets form, with a minor first peak when N5 and N6 form, and a major peak when N1-N4 form (shown in the time-distance diagrams). The intensity in all channels starts to decrease at 13:41:58~UT (or $t = 383$~s in the light-curve) after the nanojets have disappeared. This is with the exception for AIA 304, where there is continuous increase afterwards likely due to the hot Si XI emission (1.5 MK) within the 304 passband, as suggested by the temperature variation.

\section{Transverse Oscillations} \label{sec:transverse}

\begin{figure*}[ht!]
\centering
\includegraphics[width=0.8\textwidth]{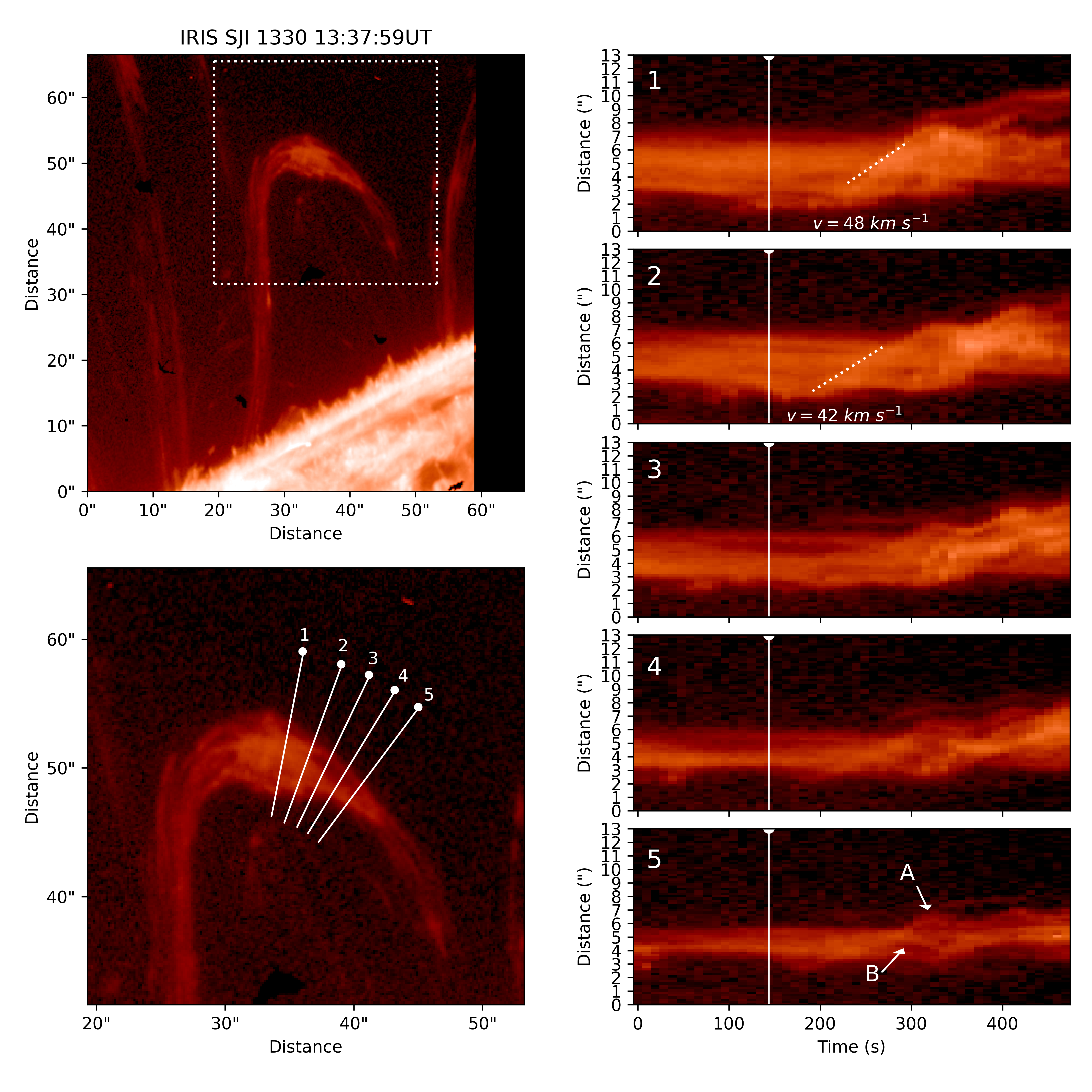}
\caption{The IRIS observation of our data taken in the SJI 1330 filtergram. The two panels in the left column show the loop-like structure of interest (top), and the zoomed-in portion from the top figure's box with slices 1-5 across the loop (bottom). The 5 slices produce the time-distance diagrams shown in the right column, where the solid white vertical lines indicate the time of the left column's snapshots. Time distances 1 and 2 show the slopes indicating the transverse velocities of the strands. After $t = 150$~s, we can see oscillating structures in the time-distance diagrams. An animation of this figure is available online, showing the evolution of the loop-like structure where the white vertical line in the time distance diagrams follows the timestamp of the panels showing the SJI images. The end of the time distance diagram is also the end of the IRIS observing period.
\label{fig3}}
\end{figure*}

After 13:37:59 UT (around $t = 150$~s in the time-distance diagrams), the nanojets that form are followed by the upward motion of several nearby rain strands, initiating a transverse motion as seen in the time-distance diagrams of Figure \ref{fig2} and Figure \ref{fig3} between t = $150-300$~s. The initial upward motion of the strands originating from the nanojets are identified by the diagonal bright slopes across the loop, where the slopes indicate velocities of $40-48$~km~s$^{-1}$ (two examples are marked in Figure \ref{fig3}). The moving strands surpass the upper edges of the loop at $t = 290$~s. After $t = 290$~s, it can be seen that the strands start an oscillatory motion while continuing to move upward at gradually slower speeds until the end  of the IRIS observation. A schematic of these dynamics is shown in the bottom panel of Figure \ref{fig2}, where the nanojets' transverse motion overshoots the resting configuration resulting in an oscillatory field motion. These oscillations have a measured period of $97 \pm 4$~s.

\begin{figure*}[ht]
\centering
\includegraphics[width=0.5
\textwidth]{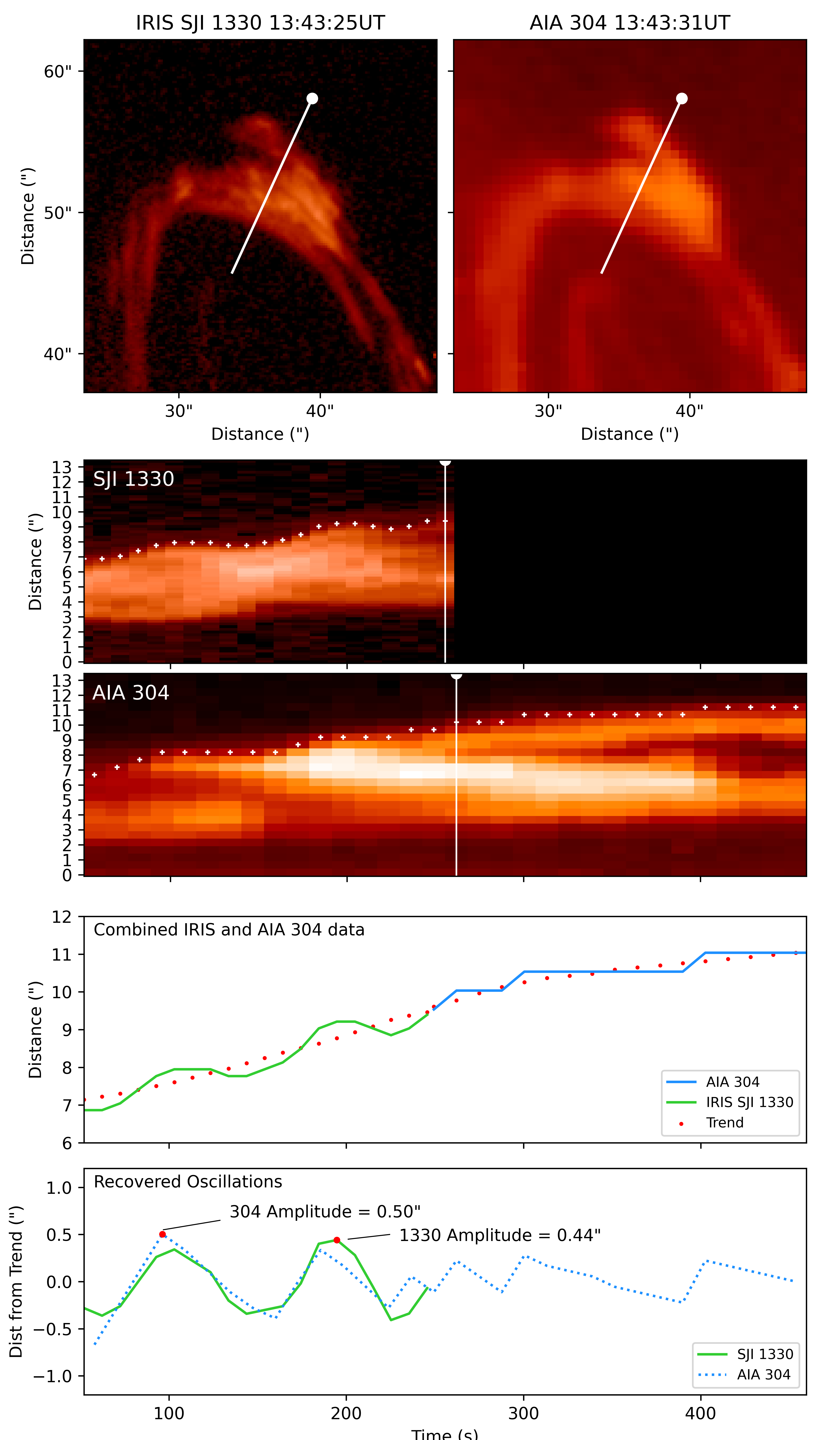}
\caption{Top row shows the SJI 1330 and AIA 304 image with the slice (white line) for the time distance diagram plots in the second and third row. The white vertical line in the time distance diagrams marks the time of the SJI and AIA image. The time-distance diagrams have the white dots marking the position of the oscillations which are selected based on an intensity threshold that separates the loop from its surrounding.  The third row shows the combined oscillation data from the SJI 1330 (first 232 seconds, in green) and AIA 304 (remaining 232 seconds, in blue) and the calculated upward moving trend (red dots). Each point in the upward moving trend is calculated by averaging a given point's value with 8 of its neighbouring points. The IRIS and AIA points are subtracted with the global trend points to recover the detrended oscillations, shown in the bottom plot. We observe 2 clear periods before it damps.
\label{fig4}}
\end{figure*}

The signatures of multiple strands oscillating can also be seen from the presence of multiple waves in Figure \ref{fig3}'s time-distance diagrams, and an example is shown in panel 5 where two waves are labeled as A and B. Note that these 2 waves are out-of-phase with each other, suggesting that each wave in the time-distance diagram comes from individual strands. Despite not being in phase with one another, all of the waves have similar periods with only 2-5 seconds differences, suggesting similar conditions across the strands. In total, we observe around $7 \pm 1$ strands that oscillates from the SJI images and time-distance diagrams. The oscillations are also most visible in the AIA 304 and 171, and faintly in 131, 193, 211, and 1600. 

We show time-distance diagrams of the oscillations in selected channels in Figures \ref{fig4} and \ref{fig5}.  Note that the sinusoidal shape of the oscillations in the AIA is not as clear as it is in the SJI, due to the small displacement and low spatial resolution of AIA. This means that it may be a challenge to identify them as oscillations had there been no accompanying observations from IRIS. 

In Figures \ref{fig4} and \ref{fig5}, there appears to be a damping of the oscillation at the extended observing time of AIA after IRIS's observing time has ended. From Figure \ref{fig4}, we can estimate the wave properties by combining the observed oscillations in IRIS and AIA to obtain the de-trended oscillations (see text/caption for further details). From the de-trended oscillations, two periods are seen with a measured maximum POS displacement ($x_{max}$) $321 \pm 30$~km. We fitted these values in a sinusoidal function of $x(t) = x_{max}\sin(\frac{2\pi t}{T})$ using the period $T$ to obtain a velocity profile of $v(t) = w\cos(2\pi t/T)$, with an amplitude $w = \frac{2\pi x_{max}}{T}$ of $21 \pm 2$~km~s$^{-1}$. The loop appears to oscillate for two periods before it damps, but it must be noted that the oscillations that we managed to recover well are only from the IRIS data thanks to its higher spatial resolution.

\section{Energy Budget of the Reconnection Event and Waves} \label{sec:energy}

As the oscillations occur at the apex of the loop, the fundamental mode is the most likely to be excited. The wave can either be a standing mode or a propagating wave. Following the nanojets, we initially observe strands that oscillate out-of-phase with each other, but eventually appear to oscillate collectively in the upper part of the loop. For this case, we can then assume that all strands oscillate with a global kink mode. However, we can also consider a multiple kink wave scenario, in which individual strands oscillate with their own kink mode (e.g. in Figure \ref{fig3}). The SJI time-distance diagrams in Figures \ref{fig2} and \ref{fig3} also show 7 strands that oscillate individually. We therefore have four possible cases: A global kink mode in which the strands oscillate as a whole, multiple kink modes guided by individual strands, and for each of these two cases we have either a standing (fundamental mode) or a propagating wave. 

In the case of a fundamental kink mode, the period $P$ of the fundamental mode is \begin{equation} \label{eq:6}
    P = \frac{2L}{c_{k}},
\end{equation}where L is the length of the loop and the phase speed $c_{k}$ is 
\begin{equation} \label{eq:7}
    c_{k} = \sqrt{\frac{\rho_{i}v_{A_i}^2 + \rho_{e}v_{A_e}^2}{\rho_{i} + \rho_{e}}}.
\end{equation} 
$c_k$ depends on the number density inside $\rho_{i}$ and outside $\rho_{e}$ the waveguide (the loop or the strand in the global kink mode or individual strands cases, respectively), and the corresponding Alfv\'en speeds $v_{A} = \frac{B}{\sqrt{\mu\rho}}$ inside and outside are written as $v_{A_i}$ and $v_{A_e}$, where the magnetic field strength B is expected to vary little under coronal conditions. The energy flux of kink modes in a bundle of loop $E_{flux}$ can be calculated from Equation \ref{eq:8} \citep{vandoorsselaere2014}, using the transverse velocity amplitude $w$ measured from the oscillations and the filling factor $f$, following 
\begin{equation} \label{eq:8}
    E_{flux} = \frac{1}{2}f(\rho_{i} + \rho_{e}) w^2 c_{k}.
\end{equation} 
The total energy can also be estimated following \cite{vandoorsselaere2014} through
\begin{equation} \label{eq:9}
    E_{total} = \pi R^2 L\left( \frac{1}{2}(\rho_{i} + \rho_{e}) w^2  - f\frac{1}{4}\rho_e \frac{c^2_k + v^2_{Ae}}{c^2_k}w^2\right) 
.
\end{equation} R is the radius of the loop (we have used half of the apparent width of the loop apex from Section \ref{sec:observations}), and $L$ is the length of the loop portion that is oscillating.

For the global kink mode and propagating wave case, the filling factor can be estimated from the area occupied by the observed number of strands in IRIS within the oscillating loop's cross-section observed in AIA \citep{vandoorsselaere2014}. Observationally, we observe 6 strands inside the loop portion (we observe 7 oscillating strands, but 1 has dampen by the time the oscillating portion forms), but it must be noted that this is a lower bound since there may be other strands that overlap one another. The oscillating loop width that contains all the oscillating strands is $3272 \pm 386$~km, whereas an individual strand has a measured width of $600 \pm 45$~km. Assuming a circular geometry, if we fill the 6 strands in the loop we will have a filling factor of $0.20 \pm 0.05$. We will also assume that the external number density (outside the loop, $\rho_e$) is $10^8$~cm$^{-3}$, and use the internal number density of the loop (surrounding the strands observed in the SJI, $\rho_i$) from the DEM analysis of $4.7 \pm 0.5 \times 10^9$~cm$^{-3}$ for this case. 

Using the values for $\rho_i$ and $\rho_e$ obtained above, the measured loop length of $120 \pm 20$~Mm in the POS for $L$, the wave period of $97 \pm 4$~s, we have $c_k = 2474 \pm 425$~km~s$^{-1}$ for the global standing kink mode. Assuming that the magnetic field inside the loop and outside the loop are similar, the estimated magnetic field is $B = 66 \pm 11$~G to match the observed period for the fundamental mode. The $E_{flux}$ and $E_{total}$ for this case are $1.2\pm 0.5$ ~$\times10^6$~erg~cm$^{-2}$~s$^{-1}$ and  $2.3 \pm 1.0 \times 10^{25}$~erg, respectively. Whereas for the global propagating mode case, we have used the measured minimum phase speed $v_{ph}$ for $c_k$ and the length of the loop portion that appears to oscillate of $16800 \pm 720$~km for $L$, to obtain $B = 32 \pm 17$~G. The  $E_{flux}$ and $E_{total}$ for this case is $0.6\pm 0.4$ ~$\times10^6$~erg~cm$^{-2}$~s$^{-1}$ and  $3.3 \pm 1.5 \times 10^{24}$~erg, respectively.

For the multiple kink mode case, the filling factor is 1 since we are resolving the strands with IRIS, and we will use the coronal rain number density of  $3.4 \pm 0.7 \times 10^{11}$~cm$^{-3}$ for $\rho_{i}$ and the DEM weighted number density for $\rho_{e}$. For the standing waves case we find that $B = 555 \pm 96$~G, and $E_{flux}$ and $E_{total}$ for a single strand are $4.3\pm1.3 \times10^8$~erg~cm$^{-2}$~s$^{-1}$ and $4.4 \pm 1.3 \times 10^{25}$~erg, respectively. We have $7 \pm 1$ strands oscillating, so the total energy released is $3.1 \pm 1.7 \times 10^{26}$~erg. For the multiple propagating kink modes case (using $v_k = c_k$), we obtain $B = 274 \pm 145$~G, and $E_{flux}$ and $E_{total}$ of $2.1\pm1.2 \times10^8$~erg~cm$^{-2}$~s$^{-1}$ and $6.1 \pm 1.5 \times 10^{24}$~erg, respectively. For 7 strands, $E_{total}$ is $4.3 \pm 2.2 \times 10^{25}$~erg.

The above are lower bound estimates since all measurements are only in the POS. Assuming that the Doppler velocity component is of the same order as the POS component, then v increases by $\sqrt{2}$. Taking account these considerations, this leads to energy flux ranges of  $1.2\pm0.4$~$\times 10^6$~erg~cm$^{-2}$~s$^{-1}$ to $8.6\pm1.3$ ~$\times 10^8$~erg~cm$^{-2}$~s$^{-1}$ from all four cases. Whereas the estimated wave energy will range between $6.5 \pm 1.5 \times 10^{24}$~erg to $6.1 \pm 2.5 \times 10^{26}$~erg.

\begin{figure*}[t!]
\centering
\includegraphics[width=0.8
\textwidth]{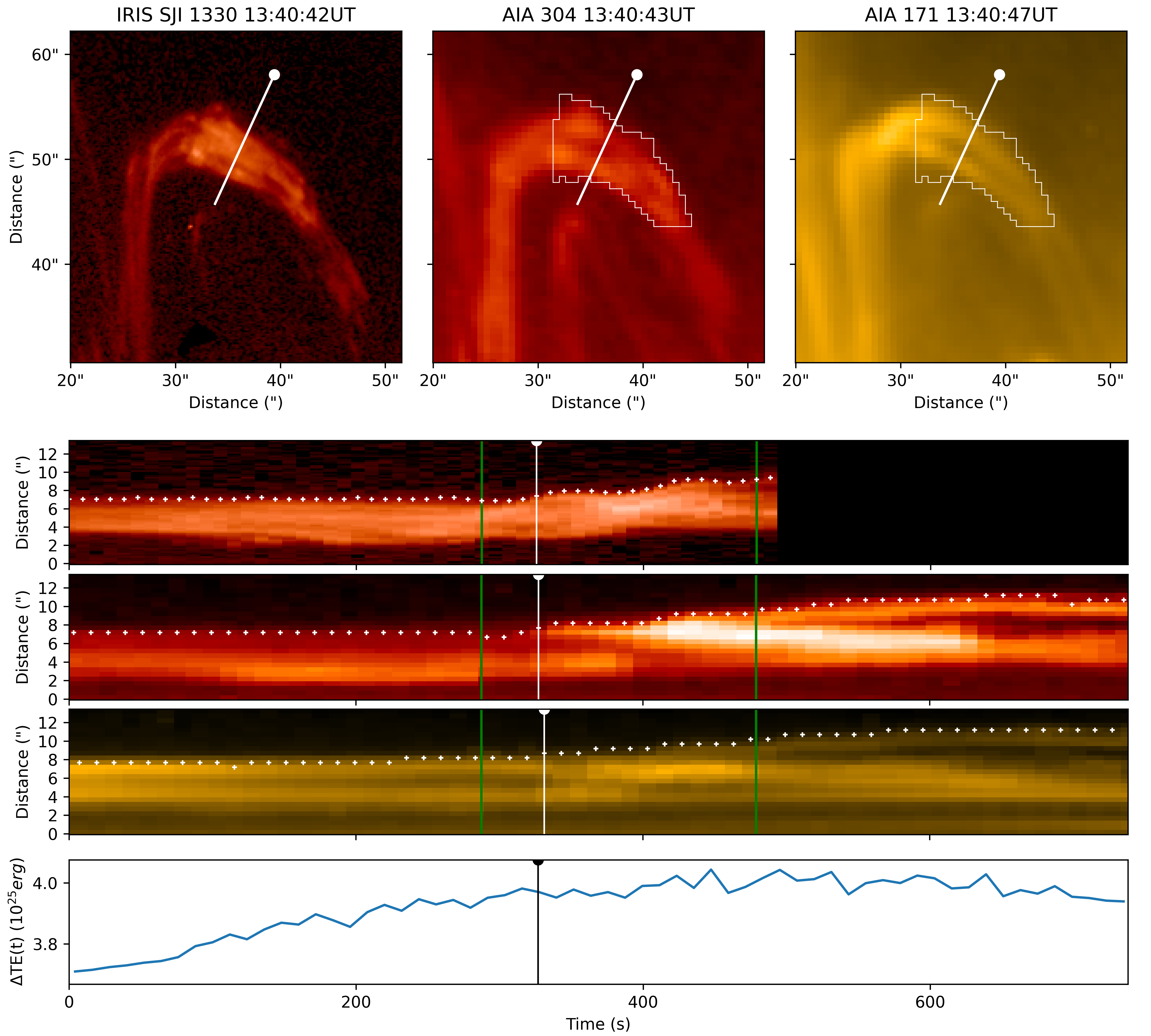}
\caption{Top row shows snapshots of the loop apex in IRIS SJI 1330, AIA 304, and AIA 171. The white line in each panel is the slice taken for the respective time-distance diagrams shown in the next three rows, where t = 0~s corresponds to 13:35:11~UT. The time-distance diagrams have the white dots marking the position of the oscillations which are selected based on an intensity threshold that separates the loop from its surrounding. The vertical white line in the time-distance diagram shows the time of the snapshots in the respective panels above. In the time-distance diagrams, the green vertical lines marks the period where we observe clear oscillations in IRIS. The plot in the bottom row shows the thermal energy release change from $t = t_0$, for the region of the loop that appears to be oscillating, estimated from the DEM for the region bounded by the white contour in the AIA images. 
\label{fig5}}
\end{figure*}

We can also calculate the total thermal energy released from the reconnection events between a given time $t_0$ (right before any nanojet occurrence) and $t$. This value $\Delta TE(t)$ can be calculated using the DEM values, for a portion of the loop that contains nanojets and is oscillating using
\begin{equation} \label{eq:10}
    \Delta TE(t) = \frac{3}{2} 2.09 V \langle n_e \rangle_{DEM} k_{B}\langle \Delta T(t)\rangle_{DEM}
\end{equation}
with the assumption of 10\% Helium abundance and a highly ionized plasma. $\langle {n}_e \rangle_{DEM}$ is the DEM weighted electron number density of the loop (from Section \ref{sec:observations}), $V = \pi R^2 L_{osc}$ is the loop portion's volume assuming a cylindrical structure and the length of the oscillating loop $L_{osc}$ of $16800 \pm 720$~km. $\langle \Delta T(t)\rangle_{DEM}$ is the average temperature variation of the loop following Equation \ref{eq:5}. This isolates the temperature change from the reconnection events associated with the nanojets that contributes to the thermal energy. We calculate the DEM for the time period starting from 13:35:11 UT - 13:41:59 UT (meaning that $t_0$ = 13:35:11 UT). Figure \ref{fig5} plots the $\Delta TE(t)$ for the loop portion, and we find that there is a continuous increase in the thermal energy to a maximum of $4.0 \times 10^{25}$~erg. 

\section{Discussions and Conclusions} \label{sec:discussion}

Our observations suggest that the transverse waves are produced by small-angle reconnection events within the structure, where the reconnection signatures can be identified by the nanojets. Prior to the nanojets, the loop did not show any oscillations as seen in the time-distance diagrams of Figures \ref{fig3} and \ref{fig5}. It is only when the nanojets form that we observe the separation of individual strands and their subsequent oscillation, which is then followed by multiple strands that eventually appear to oscillate collectively after all of the nanojets have formed. This indicates that the nanojets and the transverse MHD waves share the same generation mechanism, i.e. magnetic reconnection, and that nanojets reflect the energy available to power the oscillations. 

In the small-angle reconnection scenario, the reconnected magnetic field lines are driven sideways by magnetic tension but overshoot from their new rest position, thereby leading to transverse waves. This scenario suggests an efficient mechanism for transverse MHD wave generation. If common (as conjectured by Parker), it therefore provides an alternative explanation to the observed ubiquity of small-amplitude transverse MHD waves in the corona. 

The period of our observed oscillation is $97 \pm 4$~s with a maximum displacement and amplitude of $321 \pm 30$~km and $21 \pm 2$~km~s$^{-s}$ respectively. We have considered four cases, depending on whether the observed kink mode is a global kink mode (case in which the strands oscillate in phase on average) or a multiple kink mode (case in which each strand oscillates independently). Furthermore, we consider 2 cases for the modes: standing or propagating modes. The estimated magnetic field strengths are $32 - 66$~G for the global cases, and $274 - 555$~G for the individual strands cases. These values are considerably high but still expected from an active region producing a series of C to M-class flares (e.g. \citealt{asai2001};  \citealt{landi2021}; \citealt{wei2021}).

The produced oscillations also have similar periods suggesting similar conditions across the strands, and we observe that they occur for 2 periods before they damp. The fact that the kink waves strongly damp in a loop that is visible in the hot AIA channels throughout the event strongly suggest wave dissipation and heating. Based on the measured density, filling factor and wave properties, we estimate that the energy flux from the waves is on the order of $10^6 - 10^8$~erg~cm$^{-2}$~s$^{-1}$ for all cases, which is sufficient to balance the energy losses for active regions (\citealt{withbroe1977}). These values are lower bound estimates since we only measure projected velocities, but they indicate that braiding-induced reconnection has enough energy to power active region coronal loops. 

If the dynamics at the origin of the nanojets are what triggers the kink mode, then we may expect that the kinetic energy from the waves should match the kinetic energy of the total number of nanojets. We divide the wave's total energy released ($E_{total}$) with the average kinetic energy released by a nanojet of $7.8 \times 10^{24}$~erg to obtain an estimate of how many nanojets are required to match the wave energy for each case:  For the global standing kink mode, we only require 2-3 nanojets, which is less than the 10 clearest nanojets observed by eye. For the global propagating wave case, we need less than 1 nanojet's worth of kinetic energy.

For the multiple standing kink mode, we require around 39-40 nanojets, which is substantially more than the observed number of nanojets. We did observe around 10-12 other nanojet-like features that were too small or faint, suggesting that such numbers are indeed possible. Whereas for the multiple propagating waves, we only require around 5-6 nanojets. 

An expected feature from nanojets is the strand separation that accompanies the small-angle reconnection (\citealt{antolin2020}; \citealt{sukarmadji2022}), which would overshoot the resting field configuration. However, this strand separation may not always lead to transverse oscillations. For example, if two internal misalignments trigger nanojets that have opposite directions, the resulting oscillation may be a sausage mode rather than a kink mode. Also, if the separation is accompanied by a displacement of the footpoints then minimal or no overshoot is produced. This suggests that the reconnection events may need to have very specific conditions to produce sufficient overshoot to trigger transverse waves.

The thermal energy increase from the DEM values at the apex also shows an increase on the order of $10^{25}$~erg, with a maximum value of $4.0 \times 10^{25}$~erg just after the nanojets have stopped forming. Part of the kinetic energy of the nanojets is also likely converted into heat, and the thermal energy increase is on the same order of magnitude than the nanojet's average total energy release of $2.2 \times 10^{25}$~erg. If we assume that the thermal energy increase comes from the nanojet's total (kinetic and thermal) energy, this would mean that we only need around 2 nanojets in total. This means that only a few nanojets is required to sustain the heating seen at the apex of the loop.

We observe strands appearingly misaligned to one another, similar to the loops observed in \cite{sukarmadji2022}. Furthermore, the rain flows along the legs of the loop also appear to be misaligned, suggesting a braided structure. The entire event starts with a few nanojets, which produce transverse motion and likely create more misalignments triggering the following nanojet clusters that occur over the next five minutes. This is similar to an MHD avalanche, which is expected from previous MHD simulations of braided structures, that produce bursty nanoflare-sized heating (\citealt{hood2009}; \citealt{hood2016}; \citealt{reid2018}; \citealt{reid2020}).

The event from this work is evidence that kink waves can be a signature of braiding-induced magnetic reconnection, and that the generated kink waves can be used as a diagnostic of the energy released through reconnection.  It is likely that a large proportion of heating is still undetected through AIA: The fact that the oscillations are barely resolved in the AIA channels may wrongly suggest that there is very little wave energy in the corona. The oscillations and nanojets are only clear in IRIS, and were also only clearly detected thanks to the presence of coronal rain in the strands.

A major open question is how often the small-angle reconnection leads to kink waves, and whether a constant generation of nanojets could support the decayless kink oscillations commonly observed. If this is indeed the case, then braided field lines should be expected in oscillating loops as we require numerous misalignments to consistently produce nanojets that would sustain a decayless oscillation. However, the kink waves observed in this event damp very quickly, leading to a question of whether unresolved reconnection processes power decayless oscillations.

\begin{acknowledgments}
P. A. acknowledges funding from his STFC Ernest Rutherford Fellowship (No. ST/R004285/2). IRIS is a NASA small explorer mission developed and operated by LMSAL with mission operations executed at NASA Ames Research Center and major contributions to downlink communications funded by ESA and the Norwegian Space Centre. \textit{SDO} is part of NASA’s Living With a Star Program. All data used in this work are publicly available through the websites of the respective solar missions.
\end{acknowledgments}

\section{Appendix 1} \label{sec:appendix}

\begin{figure*}[ht]
\centering
\includegraphics[width=0.7
\textwidth]{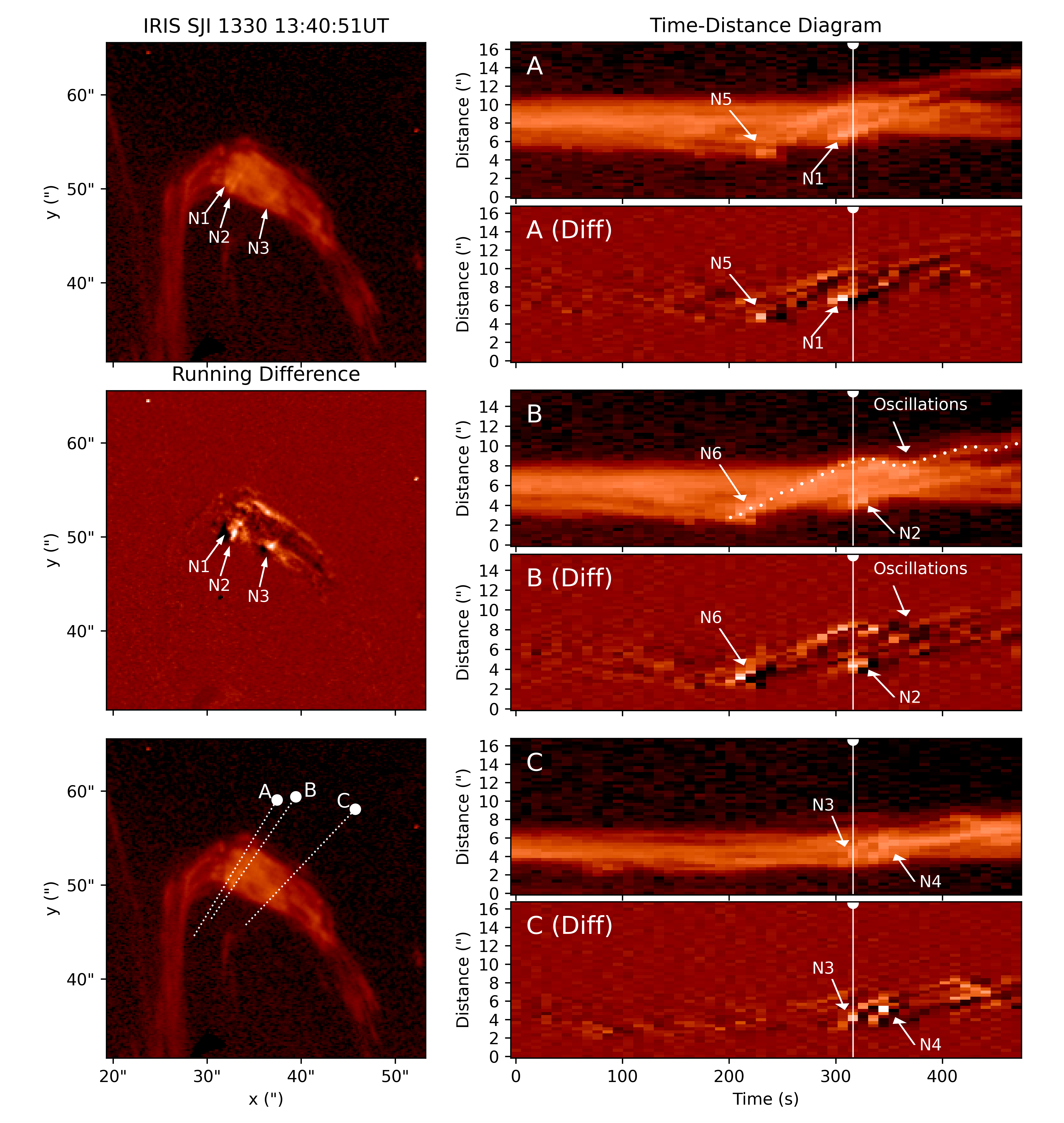}
\caption{Left column, top to bottom: The SJI snapshot with the nanojets marked, the running difference image (current snapshot subtracted with the previous snapshot), and the SJI image showing the slices A, B, and C used for the time-distance diagrams from the SJI images and the running difference images on the right column. The nanojets in the time-distance diagrams are marked with N1-N6. The white vertical lines in the time-distance diagrams mark the time of the snapshots from the left column. The causality between the nanojets and the oscillations is seen from how the strands only move upwards following a nanojet, which is then followed by an oscillation. An example of this trajectory is marked by the white dots in the time-distance diagram B, where it starts with nanojet N6, and is followed by a transverse upward motion that surpasses the upper part of the loop which then initiates the oscillatory motion at t = 300 s of the diagram. An animation of this figure is available online, showing the nanojet formation in the SJI, and the white vertical line in the time distance diagrams following the timestamp of the SJI images.
\label{fig6}}
\end{figure*}

\end{document}